\RequirePackage{fix-cm}
\documentclass[twocolumn]{svjour3_arxiv}
\usepackage{pifont,geometry,graphicx}
\usepackage[numbers,sort&compress]{natbib}
\usepackage[hidelinks,breaklinks=true]{hyperref}
\usepackage{amsmath,amssymb,amsfonts}
\usepackage{breakcites}
\usepackage{microtype}
\usepackage{gnuplottex}
\usepackage{graphicx}
\usepackage{epsfig}
\usepackage[usenames,dvipsnames]{xcolor}
\usepackage{fancyvrb}
\usepackage{stmaryrd}
\usepackage{upgreek}
\usepackage{caption}

\renewcommand{\d}{{\,\rm  d}}

\newcommand{\pd}[2]{\displaystyle\frac{\partial #1}{\partial #2}}

\newcommand{\td}[2]{\frac{{\rm d} #1}{{\rm d} #2}}
\newcommand{\pdf}[2]{\displaystyle{\partial #1}{/\partial #2}}
\newcommand{\tdf}[2]{\displaystyle{\d #1}{/\d #2}}

\newcommand{\grad}[1]{{\rm grad}\left( #1 \right)}
\renewcommand{\div}[1]{{\rm div }\left( #1 \right)}

\newcommand{\fsym}[1]{{\rm sym }( #1 )}

\newcommand{\lb}{\left(}
\newcommand{\rb}{\right)}

\newcommand{\fempty}[1]{{}}
\newcommand{\f}[1]{\mbox{$ #1 $}}
\newcommand{\topic}[1]{\\[0.8ex]{\bf  #1.}}

\newcommand{\sty}[1]{\mbox{\boldmath $#1$}}
\newcommand{\styy}[1]{{\mathbb{#1}}}
\newcommand{\fa}{\sty{ a}}
\newcommand{\fb}{\sty{ b}}

\newcommand{\fd}{\sty{ d}}
\newcommand{\fe}{\sty{ e}}

\newcommand{\fn}{\sty{ n}}

\newcommand{\ft}{\sty{ t}}
\newcommand{\fu}{\sty{ u}}

\newcommand{\fx}{\sty{ x}}

\newcommand{\fzero}{\sty{ 0}}
\newcommand{\fA}{\sty{ A}}
\newcommand{\fB}{\sty{ B}}

\newcommand{\fI}{\sty{ I}}

\newcommand{\fM}{\sty{ M}}

\newcommand{\ffC}{\styy{ C}}

\newcommand{\fsigma}{\mbox{\boldmath $\sigma$}}

\newcommand{\fxi}{\mbox{\boldmath $\xi $}}

\newcommand{\feps}{\mbox{\boldmath $\varepsilon $}}

\newcommand{\cB}{{\cal B}}

\newcommand{\cD}{{\cal D}}

\newcommand{\cP}{{\cal P}}

\newcommand{\figref}[1]{Fig.~\ref{#1}}
\newcommand{\eqreff}[1]{Eq.~\eqref{#1}}

\newcommand{\equreff}[2]{Eqs.~\eqref{#1}~and~\eqref{#2}}

\newcommand{\tabref}[1]{Table~\ref{#1}}

\newcommand{\secref}[1]{Section~\ref{#1}}

\newcommand{\EE}{\begin{equation}}
\newcommand{\Ee}{\end{equation}}
\newcommand{\FF}[1]{\begin{figure}[#1]\CT}
\newcommand{\Ff}{\end{figure}}

\newcommand{\TT}[1]{\begin{table}[#1]}
\newcommand{\Tt}{\end{table}}
\newcommand{\TC}[1]{\caption{#1.}}
\newcommand{\TB}[1]{\begin{tabular}{#1}}
\newcommand{\Tb}{\end{tabular}}
\newcommand{\II}[1]{\begin{itemize}[#1]}
\newcommand{\Ii}{\end{itemize}}
\newcommand{\EN}[1]{\begin{enumerate}[#1]}
\newcommand{\En}{\end{enumerate}}

\newcommand{\CT}{\centering}
\newcommand{\gaq}{\gamma_{\rm eq}}
\newcommand{\dgaq}{\dot\gamma_{\rm eq}}

\newcommand{\ncaq}{\nabla\zeta}
\newcommand{\ndcaq}{\nabla\dot\zeta}
\newcommand{\dcaq}{\dot\zeta}
\newcommand{\ncaqa}{|\nabla\zeta|}
\newcommand{\nggaq}{\td{\gamma}{x}}
\newcommand{\nggaqa}{|\td{\gamma}{x}|}
\newcommand{\ncaqar}{{\left(|\nabla\zeta|^2+\epsilon^2\right)}}
\newcommand\roneC[1]{\textcolor{black}{#1}}
\newcommand\rtwoC[1]{\textcolor{black}{#1}}

\newcommand{\revC}[1]{{\color{black}{#1}}}

\newcommand\rtwo{\textcolor{black}}

\newcommand{\wgc}{W_{\rm g}(\nabla\zeta)}

\makeatletter
\let\@afterindenttrue\@afterindentfalse
\makeatother
\AtBeginDocument{%
  \setlength{\oddsidemargin}{\dimexpr(\paperwidth-\textwidth)/2-1in}%
  \setlength{\evensidemargin}{\oddsidemargin}%
  \setlength{\topmargin}{%
    \dimexpr(\paperheight-\textheight)/2-\headheight-\headsep-1in}%
}
\usepackage{dblfloatfix}
\begin{document}\sloppy
\newcommand{\eqb}{\\}

\title{\rtwoC{Power-law Defect Energy in a Single-Crystal Gradient Plasticity Framework}
}
\subtitle{\rtwoC{A Computational Study}}


\author{E.~Bayerschen         \and
        T.~B\"ohlke 
}


\institute{E.~Bayerschen \and T.~B\"ohlke \at
              Institute of Engineering Mechanics (ITM), Chair for Continuum Mechanics, Karlsruhe Institute of Technology (KIT), Kaiserstr.\ 10, D-76131 Karlsruhe, Germany \\
              Tel.: +49-721-608-481-33,  +49-721-608-488-52\\
              \email{eric.bayerschen@kit.edu,}\\ thomas.boehlke@kit.edu           
}
\date{Received: date / Accepted: date}
\maketitle
 \begin{abstract}
 A \rtwoC{single-crystal} gradient plasticity model is presented that includes a \rtwo{power-law} type defect energy depending on the gradient of an equivalent plastic strain. Numerical regularization for the case of vanishing gradients is employed in the finite element discretization of the theory. Three exemplary choices of the defect energy exponent are compared in finite element simulations of elastic-plastic tricrystals under tensile loading. The influence of the power-law exponent is discussed related to the distribution of gradients and in regard to size effects. In addition, an analytical solution is presented for the single slip case \revC{supporting the numerical results}. \rtwoC{The influence of the power-law exponent is contrasted to the influence of the normalization constant.}
 \end{abstract}

\section{Introduction}
Continuum theory based material models are widely used in order to investigate the material behavior of microspecimen under different load settings like torsion (e.g., \cite{ziemann2015}). In the micron regime, non-classic material behavior like size effects occur, see, e.g., \cite{chen2015size}. These phenomena can be modeled with gradient crystal plasticity theories by incorporating an internal length scale, e.g., \cite{wulfinghoff2013gradient}. As discussed by \cite{mesarovic2010plasticity}, the need for the incorporation of an internal length scale is a consequence of the coarsening error made in the microstructural energies of continuum theories.\\
One method of incorporating internal length scales is the use of plastic strain gradient extended work hardening laws \cite{kuroda2008formulations}. Commonly, however, classic plasticity theories are extended with higher-order quantities instead. These theories can be further classified into work-conjugate and non-work-conjugate theories \cite{kuroda2006studies}. In the work-conjugate theories, the free energy is extended by a gradient related energy contribution. This defect energy considers, for example, the gradients of plastic slip or plastic strain. In the non-work-conjugate theories, however, backstresses on the individual slip systems are deduced from physical considerations of collective dislocation behavior.
Furthermore, in \cite{kuroda2008formulations}, it is shown that these two approaches are equivalent within a three-dimensional multislip context. Both types of theories can be used to predict the same kind of length scale effects in the material response.\\
The extension of the free energy is usually accomplished by adding a plastic contribution to the standard elastic free energy, see, e.g., \cite{gurtin2000plasticity}. As outlined by \cite{ortiz1999nonconvex}, an additive split of the free energy into an elastic and a plastic contribution (that can, for example, account phenomenologically for hardening in terms of the plastic slips) 
is a good first approximation for metals. In their work, the internal length scale is considered by taking into account the self energy of the dislocations (depending on the plastic part of the deformation gradient). This energy is also utilized in \cite{hurtado2013finite}, where it is shown that the occurring problem of non-differentiability in the origin of the domain of this energy can be circumvented by posing a dual formulation of the variational problem.\\
For the incorporation of the defect energy, several approaches exist. The shared aim of these is the modeling of lattice distortion due to geometrically necessary dislocations (GNDs). It is, however, also outlined in \cite{mesarovic2010plasticity} that the common association of the gradients to GNDs via the dislocation density tensor \cite{nye53} is not unique. In \cite{gurtin2000plasticity}, a geometrically non-linear theory is proposed using a defect energy being ``for convenience'' quadratic in the plastic part of the deformation gradient. A quadratic defect energy is also used in \cite{cermelli02}, formulated to depend on the geometrical dislocation density tensor within a geometrically linear theory. This formulation reduces to a quadratic form in the Burgers vector for strict plane strain. The discrete dislocation dynamics results of \cite{nicola2005effect} include a comparison of several energy formulations. Their results motivated \cite{gurtin2007gradient} to reformulate the defect energy with dependence on the densities of screw and edge dislocations rather than on the Burgers tensor. It is shown there, that this defect energy can be recast in terms of the gradient of plastic slips. The discussion of scaling regimes by \cite{scardia2014mechanics} includes this defect energy by \cite{gurtin2007gradient} in their intermediate scaling regime.\\
\roneC{A defect energy that is quadratic in the geometric dislocation density tensor has also been used in the recent variational implementations of finite gradient plasticity by \cite{miehe2014variationali,miehe2014variationalii,miehe2014variationaliii}. In \cite{miehe2014variationali}, constitutive rate-type and algorithmic incremental potentials are defined in order to formulate a general framework of inelasticity, applicable, e.g., to construct single-crystal gradient plasticity. In-line with this framework, a computational approach to gradient plasticity (of von Mises-type in the logarithmic strain space), using mixed variational principles, is proposed in \cite{miehe2014variationalii}. This includes separation into variables of long-range and short-range character, i.e., of macro-motion and micro-motion, respectively. In the variational approach \cite{miehe2014variationaliii}, such a separation is performed for multiplicative plasticity and a viscous regularization technique is proposed to overcome the problems of classic active-set search for rate-independent plasticity. In addition, mixed variational principles have been exploited for small strains in the previous work \cite{miehe2013mixed}.\\
\rtwoC{Computational aspects of gradient plasticity theories are also addressed, e.g., in the works \cite{de1992gradient,de1996some,fleck2001reformulation,pamin2003two,liebe2001theory,al2005direct,djoko2007discontinuousi,djoko2007discontinuousii,mcbride2009discontinuous,han2012plasticity,reddy2012finite,miehe2013mixed} and, more recently, in \cite{niordson2014computational,ozdemir2014modeling,Gottschalk2016443,ettehad2015numerical,kuroda2015strain}. A general methodology to incorporate the gradients of the plastic state variable within a suitable numerical scheme has been proposed in \cite{de1992gradient}. The plastic multiplier is considered as an independent variable, and the constitutive plastic equations are satisfied only in a weak sense. This has been extended using a penalty approach to allow for a $C^0$-continuous interpolation of the plastic multiplier in \cite{de1996some}. An element-free Galerkin formulation has been exploited for gradient plasticity in \cite{pamin2003two}. It ensures higher-order continuity for the shape functions to interpolate the plastic multiplier.\\
The reformulation of strain gradient plasticity presented in \cite{fleck2001reformulation} offers benefits for numerical solutions: only the primary variables and their first gradients enter the variational formulation. This is an advantage compared to earlier concepts, e.g., \cite{fleck1997strain}, where the second gradients additionally enter the variational formulation.\\
The implementation of a gradient damage framework \cite{liebe2001theory} includes the solution of the algorithmic consistency condition in weak form, and an active-set search to identify nodes in the actively damaged domain. Improved numerical costs are obtained by the implementation of gradient plasticity in \cite{al2005direct}, compared to the earlier works of, e.g., \cite{de1992gradient,de1996some}. In this approach, the additional boundary conditions need only be applied on the external boundaries. Thus, the increased effort for enforcing additional conditions on the (internal) elastic-plastic boundary has been overcome.\\
Discontinuous Galerkin formulations are explored, e.g., in \cite{djoko2007discontinuousi,djoko2007discontinuousii}, and an extension to finite strains can be found in \cite{mcbride2009discontinuous}. In the formulation \cite{reddy2012finite}, it has been shown that the return mapping algorithm for classic plasticity can also be applied to non-local plasticity such as single-crystal gradient plasticity, in an element-wise manner.\\
A numerical treatment of a strain gradient plasticity theory has been presented in \cite{niordson2014computational}. The model is of viscous type, and considers both dissipative and energetic contributions. This is in contrast to frameworks, such as the one treated in \cite{reddy2012finite}, which consider purely energetic contributions. Within the implementation \cite{ozdemir2014modeling} of the gradient plasticity theory \cite{gurtin2008theory}, all plastic slips are considered as additional degrees of freedom and interface elements are used for the grain boundary discretization. Results are presented for the two-dimensional case. In \cite{Gottschalk2016443}, however, an implementation for three dimensions is discussed and numerical examples are shown for both flow rules proposed by \cite{gurtin2008theory} as well as for an additionally proposed form.\\
The approach \cite{ettehad2015numerical} suggests using a two-step solution scheme for gradient plasticity, separating the computations into a global solution and a local approximation that facilitates meshfree methods. It is discussed that this two-step procedure simplifies the implementation of gradient theories, with an emphasis on the enforcement of boundary conditions.\\}
Recently, a variational framework with thermomechanical coupling for finite strains has been proposed \cite{bartels2015thermomechanical}. The approach \cite{anand2015stored} accounts for thermal annealing, and a computational study of this model, including viscous regularization, is presented in \cite{mcbride2015computational}. Thermal effects, however, are neglected in the present contribution as the focus is on the (mechanical part of) the defect energy.\\}
Apart from the commonly used quadratic defect energy formulations, non-quadratic forms have also been proposed in the literature. In \cite{ohno2008grain}, a defect energy is introduced that is linear in the accumulated GND-densities, thereby leading to a constant higher-order (micro) stress. A defect energy of more general type is used by \cite{gurtin2011gradient}. In their work, several special cases are discussed where they distinguish between recoverable and nonrecoverable defect energies. Recoverable defect energies in this regard means that loading and reverse loading starting from a set of slip gradients leads to the same value of the defect energy. Nonrecoverable defect energies means, however, that the same value of the defect energy is not necessarily obtained after such a loading cycle. In \cite{reddy2011role}, it is shown that the defect energy proposed by \cite{ohno2007higher} (see also \cite{kametani2012implicit}), is recoverable but non-differentiable for vanishing slip gradients. It is shown there, that the defect energy by \cite{ohno2008grain}, which is linear in the accumulated dislocation densities is, however, recoverable. In \cite{ohno2007higher}, it is pointed out that the higher-order (micro) stress resulting from their defect energy has the form of a step function. This behavior is in contrast to the resulting behavior from the quadratic defect energies which lead to gradient stresses changing linearly in dependence on, e.g., the slip gradients. Furthermore, these two different defect energy approaches are interpreted physically in \cite{ohno2007higher}. The dislocation self-energy leads to a constant gradient stress. The dislocation interaction-energy, however, leads to a gradient stress changing with slip gradients, and, thereby, accounting for the strain hardening induced by dislocation pile-ups. In \cite{hurtado2012surface}, non-local effects are accounted for in the deformation of micropillars by considering the energy of dislocation surface steps, in addition to the self-energy of dislocations. The thermo-mechanical framework of \cite{anand2015stored} also uses the linear defect energy by~\cite{ohno2007higher} in the mechanical contribution to the free energy.\\
\rtwoC{In \cite{forest2013inspection}, \rtwoC{linear} as well as \rtwoC{quadratic} defect energies are reviewed, and it is shown that the latter ones lead to physically realistic slip profiles but also to unusual scaling laws. Instead, a logarithmic defect energy is proposed and connected to the statistical theory of \cite{groma2003spatial}. Such a logarithmic defect energy is further investigated in \cite{wulfinghoff2014cyclic} and compared to a \rtwoC{linear} defect energy by analytical calculations. In addition, numerical calculations with a quadratic regularization of both energy approaches are performed to resolve the issues of differentiability for vanishing GND-densities.}\\
A non-convex strain gradient plasticity model for patterning is proposed in \cite{yalcinkaya2011deformation}. Although a quadratic defect energy is used, an additional polynomial in terms of plastic slips is employed in the free energy. The choice of the value of the internal length scale is shown to affect the distributions of plastic slip. This model is extended in \cite{yalccinkaya2012non} using a non-convex latent hardening formulation by \cite{ortiz1999nonconvex}.\\
A generalized power-law type defect energy is proposed by \cite{bardella2010size} and \cite{bardella2013latent}. It is formulated in dependence of Nye's dislocation density tensor. This dependence has originally been proposed by \cite{gurtin2002gradient}. The theory of \cite{voyiadjis2014theory} considers short-range and long-range dislocation interactions in the defect energy. With reference to \cite{bardella2010size}, it is discussed that the power exponent of the proposed defect energy governs the non-linearity of the defect energy approach. In the recent work of \cite{bardella2015modelling}, it is proposed to consider the defect energy as a function of two invariants of Nye's tensor. A power-law defect energy is proposed for this dependence and investigated for the choice of a quadratic defect energy exponent. It is also shown there that the experimental results by \cite{fleck1994strain} can be fitted with a regularized logarithmic defect energy.\\
In other recent works, it is discussed that further phenomena should be considered in the development of appropriate crystal plasticity models accounting for length scale effects. These include, for example, dislocation transport \cite{reuber2014dislocation} and vacancy diffusion in addition to coupled dislocation glide-climb mechanisms \cite{geers2014coupled}.\\
The work at hand aims at fully three-dimensional simulations with manageable computational costs. Thus, an equivalent plastic strain is used instead of considering all plastic slips as degrees of freedom. \rtwo{The theory is formulated for single-crystals using a micromorphic approach \cite{forest2009micromorphic} to include the micromorphic counterpart of the equivalent plastic strain as additional degree of freedom in the formulation.} For a brief discussion of the micromorphic modeling approach see also \cite{wulfinghoff2013gradient}. \rtwoC{The motivation to use a micromorphic quantity is the beneficiary numerical implementation of this approach within the used framework. Gradient theories can be shown to be related to micromorphic theories since they are special cases of micromorphic theories \cite{forest2003elastoviscoplastic}. This connection is constituted by constraining the micromorphic variable to be equal to its (macro) counterpart, i.e., in the work at hand, the equivalent plastic strain.\\}
The present contribution focuses on the generalization of a defect energy depending on the gradient of this micromorphic variable. This gradient may be interpreted as an approximative measure for the densities of GNDs \cite{wulfinghoff2012equivalent}. The approach of considering one equivalent plastic strain can, however, not be expected to give the same physical richness of models taking into account all plastic slips individually, see, e.g., \cite{gurtin2008theory}. \rtwo{Nevertheless, in comparison to experiments, promising results have been obtained with the use of this quantity for single-crystals \cite{ziemann2015} as well as for ensembles of crystals, e.g., oligo-crystals \cite{wulfinghoff2013gradient}. The contribution at hand is considered being a step towards a further development of this numerically efficient gradient-enhanced continuum single-crystal plasticity model.\\}
It should be noted that in the present work, it is focused on the defect energy approach and, thus, on the continuum modeling of GNDs. Nevertheless, the framework at hand would also allow one, with manageable effort, to phenomenologically include further dislocation related phenomena like, e.g., hardening effects stemming from statistically stored dislocations.
\topic{Outline} At first, the theoretical framework is presented including a \rtwoC{power-law} defect energy approach. Subsequently, the numerical implementation within a finite element setting is briefly addressed. The need for a regularization method in the case of vanishing gradients of the micromporphic variable is discussed and a numerical approach to resolve this issue is presented. Three-dimensional multislip finite element simulations of an elastic-plastic tricrystal are shown to illustrate the effect of the choice of the power-law exponent on the model behavior. Finally, an analytical single slip solution is presented and compared to the numerically obtained results.
\topic{Notation} A direct tensor notation is preferred throughout the text. Vectors and 2nd-order tensors are denoted by bold letters, e.g., by \f{\fa} or~\f{\fA}. A linear mapping of 2nd-order tensors by a 4th-order tensor is written as \f{\fA=\ffC[\fB]}. The scalar product and the dyadic product are denoted, e.g., by~\f{\fA\cdot\fB} and~\f{\fA\otimes\fB}, respectively. The composition of two 2nd-order tensors is formulated with~\f{\fA\fB}. The 2nd-order unity tensor is denoted by~\f{\fI}. Matrices are denoted by a hat, e.g., by \f{\hat\varepsilon}.
\section{Strain gradient plasticity model with a \rtwoC{power-law} defect energy approach}
\subsection{Motivation}\label{subsec:motivation}
The quality of the gradient plasticity modeling of size effect related phenomena is significantly influenced by the choice of the defect energy. Depending on the defect energy exponent \cite{bardella2010size}, it has been observed that the magnitude of size effects of the material is varying.
Common choices for the defect energy exponent are one and two. In \cite{bardella2010size}, however, the choice of a linear defect energy is argued against. This is based on the argument that a linear defect energy leads to constant, and thus, pile-up independent gradient stresses.
\rtwoC{On the path towards refining the single-crystal gradient plasticity model of \cite{wulfinghoff2013gradient}, a \rtwoC{power-law} defect energy is realized in the present contribution.} This (in general) non-linear energy is formulated in dependence of the gradient \f{\nabla\zeta} of the micromorphic counterpart~\f{\zeta} to an equivalent plastic strain~\f{\gaq}. The approach taken allows for investigations with different exponents of the defect energy.
\subsection{Basic Assumptions}
In a geometrically linear framework, the position of the material points of a body~\f{\cB} is denoted by~\f{\fx}. A Cartesian coordinate system is considered with the basis vectors reading~\f{\{\fe_1,\fe_2,\fe_3\}}. The displacement gradient is given by \f{\grad{\fu}=\pdf{u_i}{x_j}\fe_i\otimes\fe_j}, and the infinitesimal strain tensor reads \f{\feps=\fsym{\grad\fu}}. The plastic strain tensor results from a superposition of slips
\EE
\feps^{\rm p}=\sum_\alpha\lambda_\alpha\fM_\alpha^s,
\label{epsilon_plast}
\Ee
where the symmetric part of the Schmid tensor is given by~\f{\fM_\alpha^s=\fsym{\fd_\alpha\otimes\fn_\alpha}}. Slip directions are denoted by~\f{\fd_\alpha}, and~\f{\fn_\alpha} are the slip plane normals of the slip systems~\f{\alpha}. For the face-centered cubic (FCC) crystal lattice case considered here, the two possible slip directions of each slip system are considered separately. Therefore, \f{N=24}, and the positive, \rtwo{non-decreasing} quantities \f{\lambda_\alpha,\,\alpha=1,\ldots,24} are addressed by plastic slip parameters, in the following. It is remarked however, that a treatment with 12 directionally dependent plastic slips would also be suitable. Performing the common additive decomposition of the strain tensor gives the elastic strain tensor~\f{\feps^{\rm e}=\feps-\feps^{\rm p}}. An equivalent plastic strain measure is introduced as
\EE
 \gaq(\hat\lambda)=\sum_\alpha \int  \dot \lambda_\alpha  \d t = \sum_\alpha\lambda_\alpha,
\label{gammaEq}
\Ee
see, e.g., \cite{wulfinghoff2012equivalent}.

\subsection{Principle of virtual power and field equations}
The principle of virtual power is used for the derivation of the field equations. \rtwoC{A micromorphic variable~\f{\zeta} is used as an additional degree of freedom in the framework (see also the introduction of the present work).} Body forces are neglected, in the following. The virtual internal power is assumed to be given by
\EE
\delta\cP_{\rm int}=\int\limits_\cB \left(\fsigma\cdot\delta\dot\feps+\pi\delta\dcaq+\fxi\cdot\nabla\delta\dcaq\right)\d v,
\label{eq:delta_p_int}
\Ee
\rtwo{with the virtual strain rates \f{\delta\dot\feps(\delta\dot\fu)}, and the virtual rate of the micromorphic variable~\f{\delta\dcaq}.} It is remarked that \f{\feps^{\rm p}} and \f{\zeta} are a priori chosen to be independent, i.e., \f{\delta\dot\feps^{\rm p}=0} (cf. \cite{wulfinghoff2014numerically}), \rtwo{and that the micromorphic variable~\f{\zeta} is penalized to be equal to its (macro) counterpart, the equivalent plastic strain~\f{\gaq}, see \eqreff{eq:stored_energy} and \eqreff{eq:wchi}}.
The Cauchy stress tensor is denoted by~\f{\fsigma}, the stress associated to the rate~\f{\dcaq} is denoted by \f{\pi}, and the gradient stress~\f{\fxi} is associated to the rate~\f{\ndcaq}. All three occurring stress quantities, \f{\fsigma}, \f{\pi}, and \f{\fxi}, are work conjugate to their associated kinematic quantities \f{\dot\feps}, \f{\dcaq} and \f{\ndcaq}, respectively. It is assumed that~\f{\zeta} is a continuous quantity. Consequently, possible jumps of~\f{\zeta}, e.g., at grain boundaries in the body~\f{\cB}, are neglected. This assumption is not expected to be transferable to plastic slips of individual slip systems, in general. In the context of an overall description of plastic slip, it is remarked that the calculations of an effective plastic strain in the experimental work of \cite{abuzaid2012slip} leaded to continuous distributions of this overall quantity across many GBs. Thus, it appears to be reasonable to consider~\f{\zeta} as a continuous quantity in a first approach.\\
Furthermore, the virtual power of external forces is assumed to have the contributions
\EE
\delta\cP_{\rm ext}=\int\limits_{\partial\cB_{\rm t}}\bar\ft\cdot\delta\dot\fu\d a+\int\limits_{\partial\cB_{\Xi}}\bar\Xi\,\delta\dcaq\d a,
\label{eq:delta_p_ext}
\Ee
where power is expended at the external boundaries~\f{\partial\cB_{\rm t}\cup\partial\cB_{\Xi}=\partial\cB} by the tractions~\f{\bar\ft}, and the microtractions~\f{\bar\Xi}, respectively. In \eqreff{eq:delta_p_ext}, it has been exploited, already, that the virtual rates \f{\{\delta\dot\fu,\delta\dot\zeta\}} vanish for given~\f{\{\fu,\zeta\}} at the Dirichlet boundaries \f{\partial\cB_{\rm u}}. By using the principle of virtual power, i.e., by letting the virtual power of the internal forces be equal the virtual power of the external forces, \f{\delta\cP_{\rm int}=\delta\cP_{\rm ext}}, and applying Gauss' theorem, the field equations and the Neumann boundary conditions can be derived (cf.\ appendix~\ref{app:field_equations}). They are shown in Box~1.
\begin{figure*}[hbtp]
\caption*{Box 1: Field equations and boundary conditions.}
\centering{
\renewcommand{\arraystretch}{1.1}
\TB{|lrll|}
\hline
Linear momentum balance   & \f{\fzero}        & \f{=\div{\fsigma}}&\f{\quad\forall\fx\in\cB}\\
  Microforce balance& \f{\pi}&\f{=\div{\fxi}}&\f{\quad\forall\fx\in\cB}\\
  Neumann BCs for: Cauchy stress            & \f{\fsigma \fn}     & \f{= \bar\ft} & \f{\quad\mathrm{on}\ \partial \cB_t}  \\
		\phantom{Neumann BCs for: }Gradient stress &\f{\fxi \cdot \fn}	&\f{= \bar \Xi} &\f{\quad\mathrm{on} \ \partial \cB_{\,\Xi}}
		\\\hline
\Tb
}
\end{figure*}
The classic quasistatic balance of linear momentum is supplemented by an additional microforce balance taking into account the gradient stress arising from the gradient extension of the free energy. Consequently, in addition to the Neumann boundary conditions (BCs) for the Cauchy stress, Neumann BCs for the gradient stress are necessary, as well. It is remarked that contrary to the theories using all plastic slips as additional degrees of freedom (DOF), \rtwoC{in the theory at hand, only one scalar, the micromorphic variable~\f{\zeta} (penalized to be equal to the equivalent plastic strain~\f{\gaq}), is employed as additional DOF.} Consequently, the gradient stress~\f{\fxi}, associated to the gradient of the micromorphic counterpart of the equivalent plastic strain, is identical on all slip systems. In theories considering all plastic slips as individual degrees of freedom, however, different gradient stresses are obtained on the individual slip systems. \rtwoC{It should be noted that microhard conditions are employed on the interfaces between the crystals (i.e., on the grain boundaries) and on the external boundaries in the lateral direction (i.e., the two planar surfaces with minimum and, respectively, maximum $x$-coordinate). This means that the equivalent plastic strain remains zero, there.}
\subsection{Constitutive assumptions}
\subsubsection{Free energy density}
The free energy density is assumed to have the form
\EE
\begin{split}
W(\feps,\hat\lambda,\zeta,\nabla\zeta)=W_{\rm e}(\feps,\feps^{\rm p}(\hat\lambda))+W_{\rm g}(\nabla\zeta)\eqb+W_{\chi}(\zeta-\gamma_ {\rm eq}(\hat\lambda)),
\end{split}
\label{eq:stored_energy}
\Ee
with a quadratic elastic contribution
\begin{equation}
 W_{\rm e}(\feps,\feps^{\rm p}(\hat\lambda))=\frac{1}{2}(\feps-\feps^{\rm p})\cdot\ffC[\feps-\feps^{\rm p}],
\end{equation}
where~\f{\ffC} denotes the elastic stiffness tensor, and a generally non-quadratic gradient-related defect energy contribution
\newcommand{\wg}{W_{\rm g}(\nabla\zeta)}
\EE
\wg=W_0\left(\frac{\ncaqa}{g_0}\right)^m.
\label{eq:gradgammam}
\Ee
Here, \f{W_0} is an initial defect energy. For the case of a defect energy being linear in the dislocation density, this value is commonly taken to be \f{W_0=\bar\alpha G}, cf.\ \cite{ohno2007higher,hirth1982theory}. The shear modulus is denoted by~\f{G}, and~\f{\bar\alpha\approx0.5\ldots1} is often assumed to be a constant. The quantity~\f{g_0} denotes a normalization constant and can, thus, be related to the internal length scale that is introduced by the defect energy in the theory. In the case of a linear defect energy, this constant could be taken to be the inverse of the Burgers vector length, i.e., \f{g_0=1/b} (see, e.g., \cite{ohno2007higher}).
It should be noted that the gradient extension is performed here in terms of the micromorphic field variable~\f{\zeta} instead of~\f{\gaq}, cf.\ \eqreff{eq:stored_energy}. The coupling of both quantities is accomplished with the energy density
\EE
\label{eq:wchi}
W_\chi(\zeta-\gamma_ {\rm eq}(\hat\lambda)) = \frac{H_\chi}{2}(\zeta-\gamma_{\rm eq})^2,
\Ee
which penalizes deviations of~\f{\zeta} from~\f{\gamma_{\rm eq}}. By use of a large penalty factor~\f{H_\chi} it is achieved that \f{\zeta\approx\gaq} (see also \cite{wulfinghoff2013gradient}).
\subsubsection{Dissipation}
The total dissipation is the difference between the power of the external forces and the rate of the free energy in the bulk,
\EE
D_{\rm tot}= \int \limits_\cB \cD \d v  = \cP_{\rm ext} - \int \limits_\cB \dot W \d v  \geq 0,
\label{eq:dissipation_total}
\Ee
if thermal and grain boundary effects are neglected. By letting \f{\cP_{\rm ext}=\cP_{\rm int}=\int_\cB \left(\fsigma\cdot\dot\feps+\pi\dcaq+\fxi\cdot\nabla\dcaq\right)\d v}, and substituting~\eqreff{eq:stored_energy} in~\eqreff{eq:dissipation_total}, the dissipation~\f{\cD} can be reformulated as 
 \EE
\label{eq:bulk_dissipation}
\begin{split}
 \cD = \left(\fsigma - \pd{W_{\rm e}}{\feps}\right) \cdot \dot \feps - \pd{W_{\rm e}}{ \feps^{\rm p}}   \cdot \dot \feps^{\rm p}- \pd{W_\chi}{\gaq}  \dgaq\eqb+\left(\pi- \pd{W_\chi}{\zeta}\right) \dcaq + \left(\fxi -\pd{W_{\rm g}}{\nabla \zeta} \right) \cdot \nabla \dcaq \geq0.
\end{split}
\Ee
Introducing the abbreviation $\check p=\pdf{W_\chi}{\gaq}=- \pdf{W_\chi}{\zeta}$, assuming the stresses~\f{\fsigma}, \f{\pi} and~\f{\fxi} to be purely energetic, and substituting~\f{\pdf{W_{\rm e}}{\feps^{\rm p}}=-\fsigma}, leads to the reduced dissipation inequality
\EE
 \cD = \fsigma \cdot \dot \feps^{\rm p} - \check p\, \dot \gamma_{\rm eq}  \geq 0.
\label{eq:diss_reduced}
\Ee
Substituting \equreff{epsilon_plast}{gammaEq} in~\eqreff{eq:diss_reduced} gives
\EE
\cD= \sum_\alpha (\tau_\alpha - \check p)\, \dot \lambda_\alpha\geq0,
\Ee
with the resolved shear stresses~\f{\tau_\alpha=\fsigma\cdot\fM_\alpha^s}.
Furthermore, based on \cite{cermelli02}, it is assumed that the bulk dissipation is induced by the slip system contributions, i.e., by the dissipative shear stresses~\f{\tau_\alpha^{\rm d}} and the slip parameter rates~\f{\dot\lambda_\alpha},
\EE
\cD=\sum_\alpha\tau_\alpha^{\rm d}\dot\lambda_\alpha.
\Ee 
Consequently, the dissipative shear stresses can be expressed by
\EE
\tau_\alpha^{\rm d}={\tau_\alpha}-\check p.
\Ee
Finally, using~\f{\check p = -\pi} from~\eqreff{eq:bulk_dissipation} and the microforce balance (\f{\pi=\div\fxi}, see Box~1), gives
\EE
\label{eq:tau_xi}
\tau_\alpha^{\rm d}=\tau_\alpha+\div\fxi.
\Ee
%
\subsubsection{Flow rule}
The flow rule is assumed to be of an overstress type, formulated in the plastic slip parameter rates~\f{\dot\lambda_\alpha},
\EE
\label{eq:powerLaw}
\begin{split}
 \dot \lambda_\alpha &= \dot \gamma_0 \left\langle
\frac{\tau_\alpha^{\rm d} - \tau^{\rm C}_0}{\tau^{\rm D}}
\right\rangle^p
\eqb&=\dot \gamma_0 \left\langle
\frac{\tau_\alpha+\div\fxi- \tau^{\rm C}_0 }{\tau^{\rm D}}
\right\rangle^p.
\end{split}
\Ee
Here, \f{\dot\gamma_0} is the reference shear rate, \f{\tau^{\rm D}} is the drag stress, and~\f{p} is the rate sensitivity exponent. The initial yield stress of the slip systems is denoted by~\f{\tau_0^{\rm C}}. It is remarked that on all slip systems~\f{\alpha}, the term~\f{\div\fxi} is acting identically. This is due to the usage of the gradient of the micromorphic variable \f{\nabla\zeta} in the defect energy~\eqreff{eq:gradgammam}, instead of considering the gradients of the slips of all slip systems individually.

\section{Finite Element Implementation}
\subsection{Linearization of the variational form}
The theory is implemented with finite elements (FE) in an in-house FE-code. An enhanced time-integration algorithm is used that allows for large time-steps compared to numerical standard schemes (see \cite{wulfinghoff2013equivalent}). In order to discuss the key concepts in the implementation of the \rtwoC{power-law} defect energy approach, \eqreff{eq:gradgammam}, the linearization of the principle of virtual power is addressed, briefly.
The linearization of \equreff{eq:delta_p_int}{eq:delta_p_ext} in~\f{\delta \dot \fu} yields
\begin{equation}
\begin{split}
\int\limits_{\cB}\left(\pd{\fsigma}{\feps}[\Delta\feps]+\pd{\fsigma}{\zeta}\Delta\zeta\right)\cdot\delta \dot \feps\d v=\eqb-\int\limits_{\cB}\fsigma\cdot\delta \dot \feps\d v+\int\limits_{\partial\cB_{\rm t}}\bar\ft\cdot\delta \dot \fu\d a\quad\forall\delta \dot \fu .
\end{split}
\label{eq:disc_1}
\end{equation}
Furthermore, the linearization of \eqreff{eq:delta_p_int}, and \eqreff{eq:delta_p_ext} in~\f{\delta \dot \zeta}, considering additionally~\f{\pi=-\check p}, gives
\begin{equation}
\begin{split}
&\int\limits_{\cB}\bigg(\delta \dot \zeta\Big(-\pd{\check p}{\zeta}\Big)\Delta\zeta
-\delta \dot \zeta\pd{\check p}{\feps	}\cdot\Delta\feps\eqb&+\pd{\fxi}{\nabla\zeta}[\nabla(\Delta\zeta)]\cdot\nabla(\delta \dot \zeta)\bigg)\d v
=\eqb&-\int\limits_{\cB}\bigg(-\check p\,\delta \dot \zeta+\fxi\cdot\nabla(\delta \dot \zeta)\bigg)\d v+\int\limits_{\partial\cB_\Xi}\bar\Xi\,\delta \dot \zeta\d a.
\end{split}
\label{eq:disc_2}
\end{equation}
The numerical integration of \eqreff{eq:disc_1} and \eqreff{eq:disc_2} is outlined in \cite{wulfinghoff2013equivalent} in a more general formulation, including grain boundary contributions.\\
Obviously, the FE implementation of \eqreff{eq:disc_1} and \eqreff{eq:disc_2} requires the computation of the gradient stress~\f{\fxi=\pdf{W_{\rm g}(\nabla\zeta)}{\nabla\zeta}} and its derivative~\f{\pdf{\fxi}{\nabla\zeta}}. Using the \rtwoC{power-law} defect energy formulation, \eqreff{eq:gradgammam}, these quantities can be expressed by

\EE
\fxi=\pd{\wgc}{\ncaq}=m\frac{W_0}{g_0^m}\ncaqa^{m-2}\ncaq,
\label{eq:xi_nonreg}
\Ee
and
\EE
\begin{split}
\pd{\fxi}{\ncaq}=&m\frac{W_0}{g_0^m}\ncaqa^{m-2}\fI\eqb&+m(m-2)\frac{W_0}{g_0^m}\ncaqa^{m-4}\ncaq\otimes\ncaq,
\end{split}
\label{eq:pdxi_nonreg}
\Ee
respectively. From~\eqreff{eq:xi_nonreg} and~\eqreff{eq:pdxi_nonreg} it becomes apparent that for an arbitrary power exponent~\f{m}, the gradient stress in this formulation is neither defined nor differentiable in the case of vanishing gradients~\f{\nabla\zeta}, i.e., for~\f{|\nabla\zeta|\rightarrow0}. However, for the special case of a quadratic defect energy with \f{m=2}, the gradient stress reads
\EE
\fxi=2\frac{W_0}{g_0^2}\ncaq,
\Ee
and its derivative with respect to~\f{\nabla\zeta} is
\EE
\pd{\fxi}{\ncaq}=2\frac{W_0}{g_0^2}\fI.
\Ee
Consequently, for this special case, the gradient stress is defined and differentiable, even for vanishing gradients~\f{\nabla\zeta}. It is remarked that the quadratic defect energy leads to a gradient stress~\f{\fxi} that is linear in the gradient~\f{\nabla\zeta}. Choosing \f{m=1}, however, yields a constant gradient stress~\f{\fxi}.\\
In order to use the \rtwoC{power-law} defect energy with varying exponents \f{m}, a regularization of the \rtwoC{power-law} defect energy, \eqreff{eq:gradgammam}, for vanishing gradients~\f{\nabla\zeta}, is outlined in the following.
\subsection{Regularization of the \rtwoC{power-law} defect energy approach}
\rtwo{A standard regularization approach is undertaken here, e.g., \cite{glowinski1981}.} Regularized defect energy associated quantities are denoted by a~\f{\tilde{(\cdot)}}, in the following. The regularized defect energy is introduced as
\newcommand{\wgt}{\tilde W_{\rm g}(\nabla\zeta)}
\EE
\begin{split}
\wgt&=W_0\left(\frac{{\ncaqar}^{\frac{1}{2}}}{g_0}\right)^m\eqb&=\frac{W_0}{g_0^m}\ncaqar^{\frac{m}{2}}.
\end{split}
\label{eq:gammam_reg}
\Ee
The numerical parameter~\f{\epsilon} is chosen sufficiently small but non-zero (see \secref{sec:convergence}), and leads to finite values of the derivatives of \eqreff{eq:gammam_reg}, even in the case of vanishing gradients \f{\nabla\zeta}.\\
The gradient stress corresponding to the employed regularization is
\newcommand{\xit}{\tilde \fxi}
\EE
\xit=\pd{\wgt}{\ncaq}=m\frac{W_0}{g_0^m}\ncaqar^{\frac{m}{2}-1}\ncaq,
\Ee
and its derivative with respect to the gradient is given by
\EE
\begin{split}
&\pd{\xit}{\ncaq}=m\frac{W_0}{g_0^m}\ncaqar^{\frac{m}{2}-1}\fI\eqb&+m(m-2)\frac{W_0}{g_0^m}\ncaqar^{\frac{m}{2}-2}\ncaq\otimes\ncaq.
\end{split}
\Ee
For the special case~\f{m=2}, this regularized formulation recovers the non-regularized defect energy, i.e., \f{\xit=2{W_0}\ncaq/{g_0^2}}, and \f{\pdf{\xit}{\ncaq}=2{W_0}\fI/{g_0^2}}, respectively.
%
%
If the regularization parameter~\f{\epsilon} is set to vanish, the original formulation, \eqreff{eq:gradgammam}, of the \rtwoC{power-law} defect energy is recovered, i.e., \f{\xit=\fxi}, and \f{\pdf{\xit}{\ncaq}=\pdf{\fxi}{\ncaq}}, respectively.
%
\subsection{Numerical time integration and algorithmic tangents}
The numerical time integration requires the determination of the stresses in~\equreff{eq:disc_1}{eq:disc_2} and the computation of the algorithmic tangents. These computations follow along the scheme discussed in~\cite{wulfinghoff2013gradient}, but the grain boundary contributions are neglected \rtwoC{since only microhard conditions are considered on the grain boundaries throughout the present work.} Furthermore, isotropic hardening of Voce-type is not considered in the work at hand, and thus, \f{\check p} does not have an isotropic hardening contribution~\f{\beta} as in the above mentioned work. An implicit Euler scheme is applied to~\eqreff{eq:powerLaw}. For details on this it is referred to~\cite{wulfinghoff2013equivalent}.
\section{Numerical results}\label{sec:num}
\subsection{Material model, geometry, and boundary conditions}
\begin{figure}[htbp]
\centering
\includegraphics[width=0.85\linewidth]{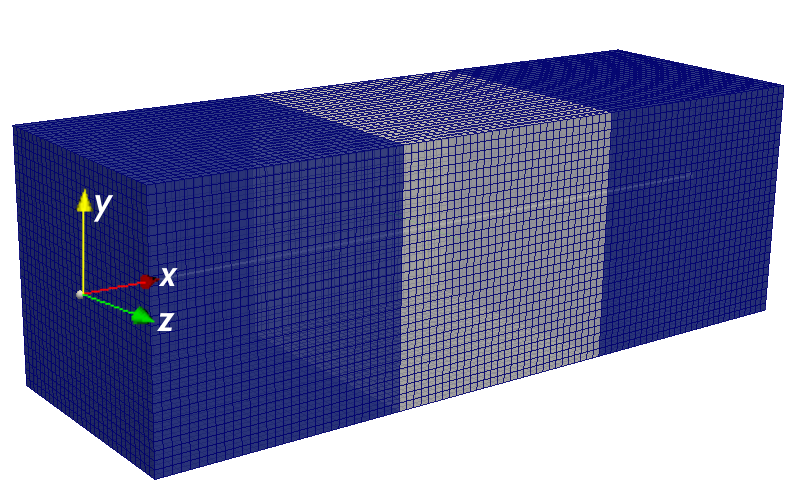}
\caption{Spatial discretization of tricrystal with elastic bounding grains (blue) and line segment along the central axis (white).}
\label{fig:trikristall}
\end{figure}

The material considered throughout the finite element simulations in this work is, for simplicity, chosen to be elastically isotropic with material constants similar to aluminum, see~\tabref{tab:paramer_general}.
 \begin{table}[htbp]
\renewcommand{\arraystretch}{1.5}
\TC{Model parameters for tensile test simulations of aluminum-like tricrystals}
\label{tab:paramer_general}
\scriptsize
\centering
\begin{tabular}{||ccccc||}
\hline
\f{G}&\f{\nu}&\f{W_0}&\f{\bar\alpha}&\f{p}\\\hline
25 GPa&0.3&\f{\bar\alpha G}&$0.5$&20\\\hline\hline
\f{\dot\gamma_0}&\f{\tau_0^{\rm C}}&\f{\tau^{\rm D}}&\f{\epsilon}&\f{H_\chi}\\\hline
10$^{-3}$ 1/s&33.5 MPa&1 MPa&10\f{^{-6}}&10\f{^8} MPa\\\hline
\end{tabular}
\end{table}
Fully anisotropic plasticity, however, is accounted for in a multislip setting for FCC materials. A tricrystal composed of cubic grains (see \figref{fig:trikristall}) is simulated to demonstrate the behavior of the model for different defect energy exponents~\f{m}. The same initial defect energy~\f{W_0} is used in all simulations. At first, the three grains are considered to be in a standard crystal lattice orientation (i.e., they are oriented $\langle 100 \rangle$ with respect to the $x$-axis of the Cartesian~\f{x}-\f{y}-\f{z}-system).
In order to allow an assessment of the \rtwoC{power-law} defect energy approach, the behavior of the occurring gradients of the micromorphic variable~\f{\zeta} close to the grain boundaries is isolated by setting the two bounding grains to behave purely elastic. Subsequently, a fully elastic-plastic tricrystal is discussed, additionally, where the two grain boundaries do not allow any dislocation transmission across, i.e., they are treated as microhard boundaries. In the theory at hand, these microhard boundaries can be included as a subset of an arbitrarily chosen grain. No additional grain boundary terms need, therefore, to be considered in the theoretical model (cf.\ also the discussion on this choice in~\cite{wulfinghoff2013equivalent}).\\
The tricrystal of dimensions \f{3\times3\times9\,\upmu}m (see~\figref{fig:trikristall}) is loaded strain controlled up to an overall strain of~\f{\varepsilon=0.005} under tensile loading along the $x$-axis. Lateral contraction is allowed for by the boundary conditions on the two bounding planes at~\f{x=0\,\upmu}m, and~\f{x=9\,\upmu}m. In the following, distributions of the micromorphic field variable~\f{\zeta} are investigated for the different elast-plastic cases under consideration. The line segment highlighted in \figref{fig:trikristall} is, therefore, utilized to obtain plots of the field distribution.
\subsection{Convergence of results}\label{sec:convergence}
\begin{figure}[h]
\centering
\includegraphics[width=0.85\linewidth]{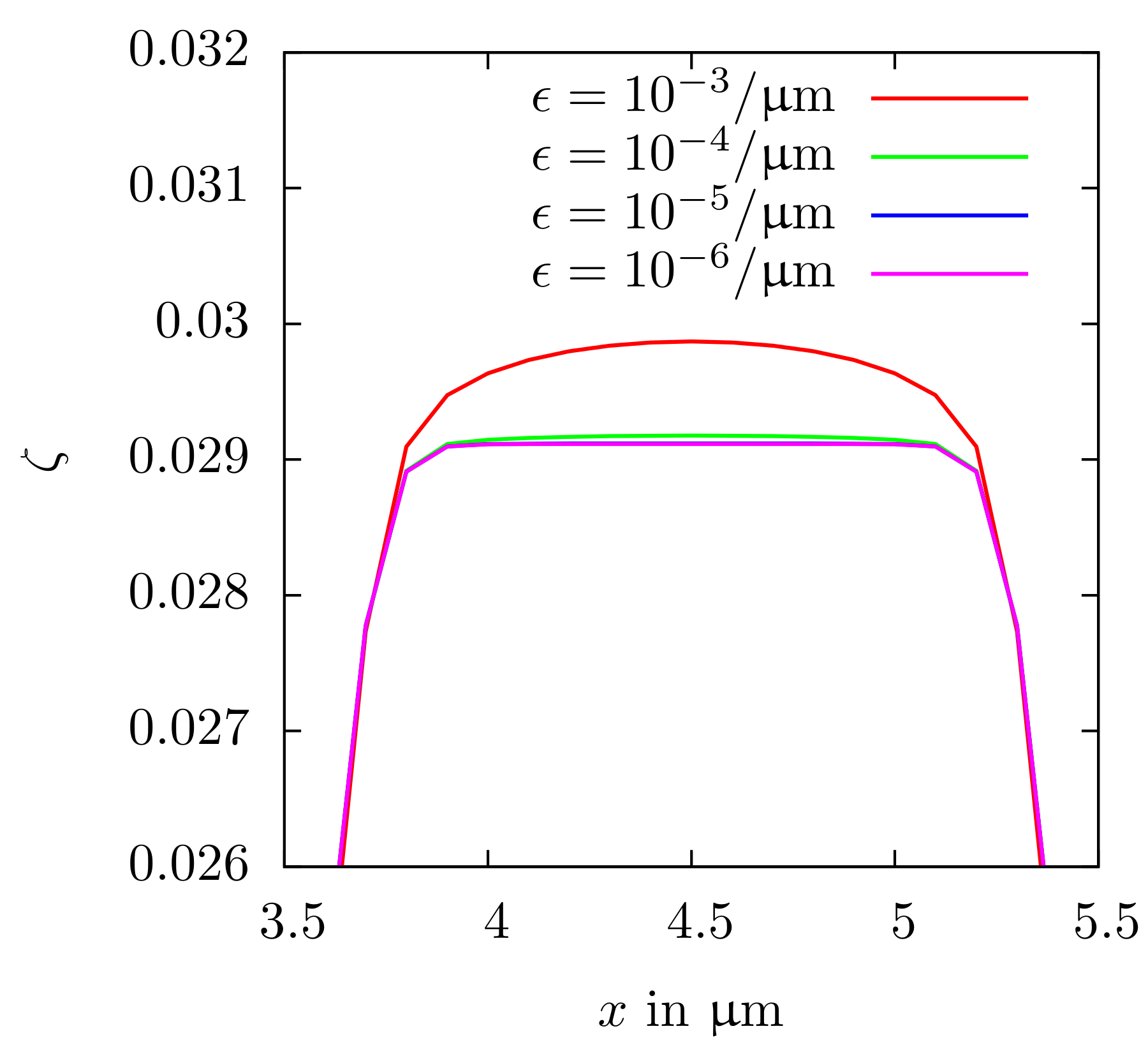}
\caption{Distribution of micromorphic variable \f{\zeta} in central grain of tricrystal with elastic bounding grains for discretization with 349804~DOF, and different numerical regularization parameters~\f{\epsilon}.}
\label{fig:final_stress_convergence_02}
\end{figure}

The convergence of the numerical results is discussed twofold, here.
\begin{figure}[h]
\centering
\includegraphics[width=0.85\linewidth]{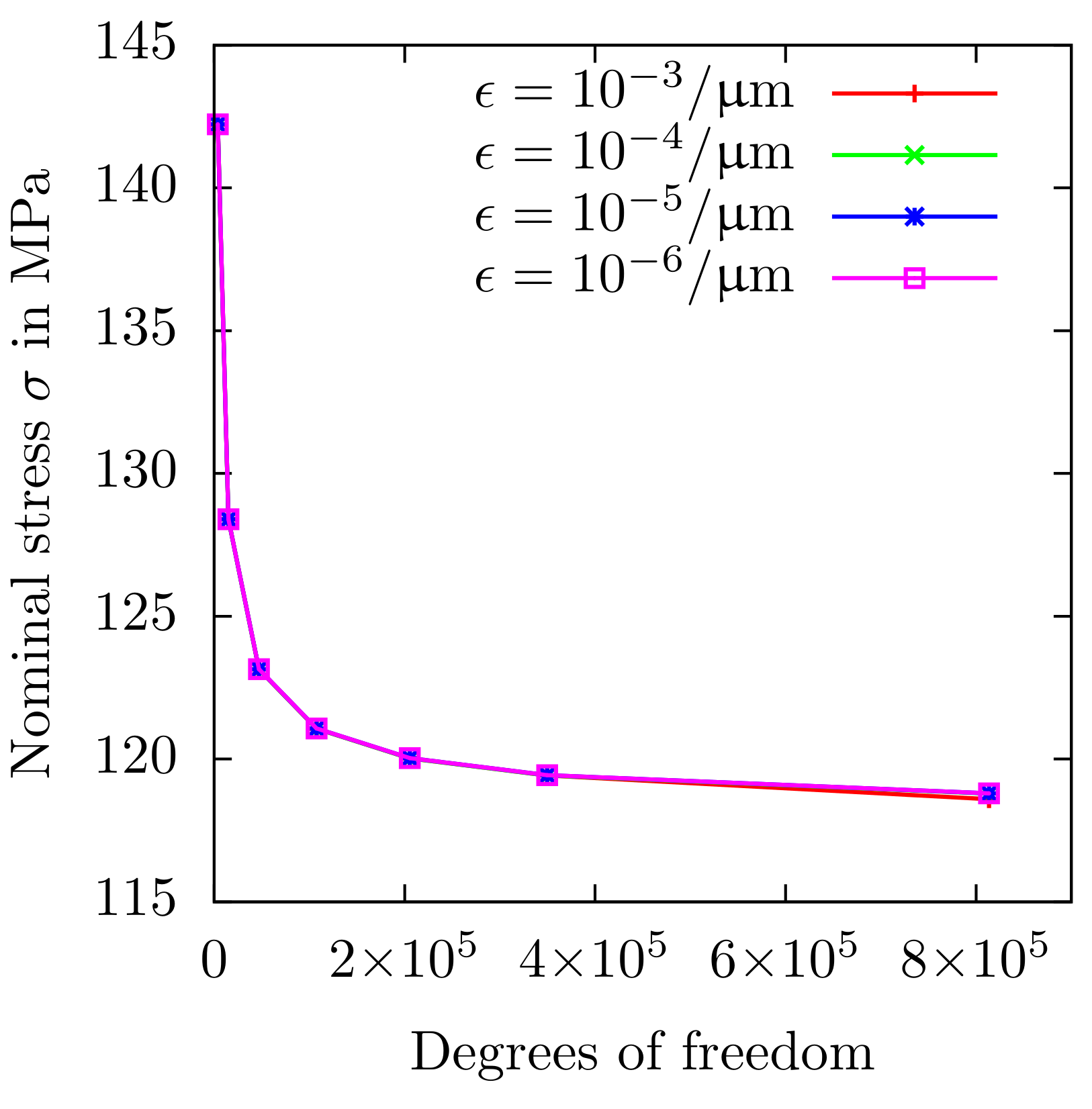}
\caption{Final nominal stress at $t=1.0$~s vs degrees of freedom for tricrystal with elastic bounding grains, different spatial discretizations, and different numerical regularization parameters~\f{\epsilon}.}
\label{fig:final_stress_convergence_01}
\end{figure}
Besides convergence in regard to the spatial discretization of the finite element mesh, the convergence with respect to the regularization parameter~\f{\epsilon} is investigated. The defect energy exponent used in this section is $m=1.1$, and the normalization constant is $g_0=450.21$/\f{\upmu}m.
As can be seen in~\figref{fig:final_stress_convergence_01}, the regularization parameter~\f{\epsilon} is of negligible influence on the final nominal stress at the end of the loading. It also does not influence the convergence behavior regarding the spatial discretization.
In~\figref{fig:final_stress_convergence_02}, it is shown that a sufficiently small choice of the regularization parameter leads to coinciding distributions of~\f{\zeta} along the $x$-axis of the central grain for a further decrease in~\f{\epsilon}.
\subsection{Results for the defect energy with different exponents $m$}
\subsubsection{Tricrystal with elastic bounding grains}
In order to investigate the model behavior for different exponents~\f{m} in the defect energy approach, \eqreff{eq:gammam_reg}, two approaches are undertaken. At first, numerical results are obtained with a subsequent comparison to results from an analytical solution, see~\secref{sec:anal}.
\begin{figure*}[t]
\centering
\includegraphics[width=\linewidth]{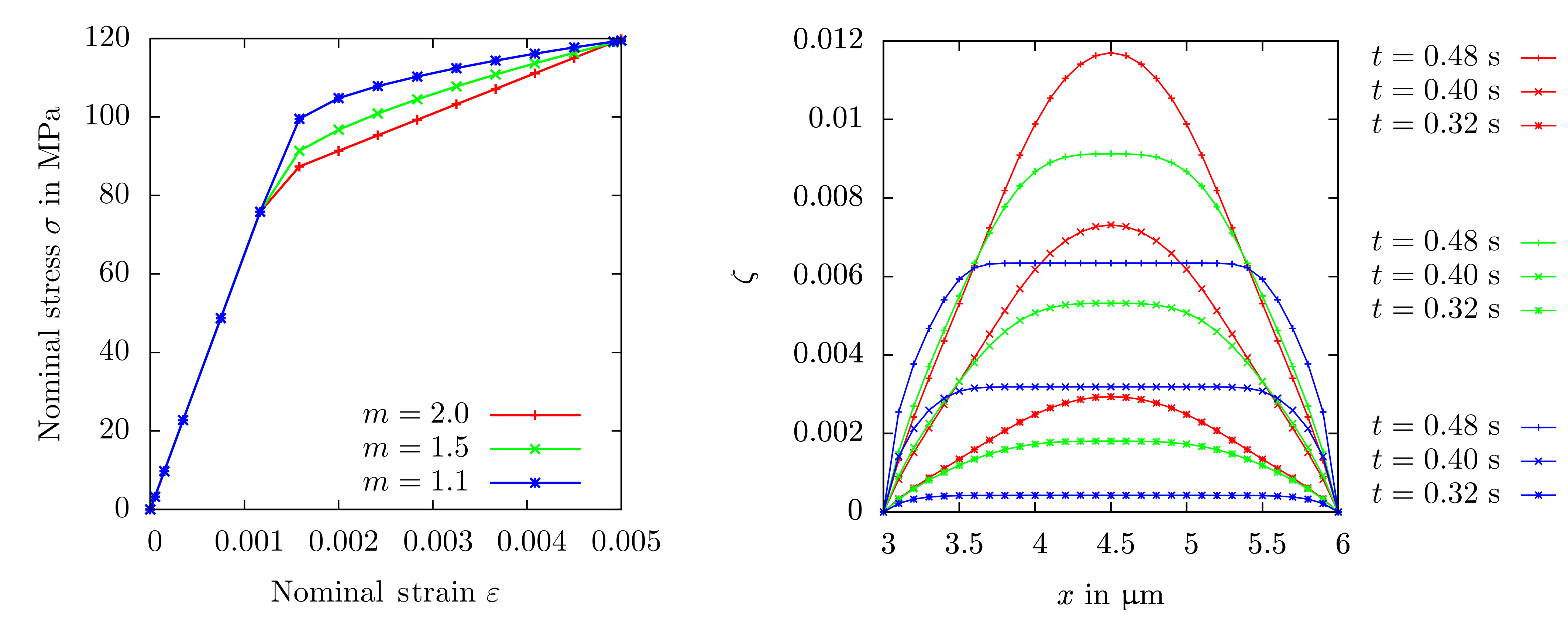}
\caption{Left: Stress-strain curves for tricrystal with elastic bounding grains and different defect energy exponents~\f{m}. Right: $\zeta$-distributions in the central grain along the $x$-axis for three time steps.}
\label{fig:zeta_vs_m}
\end{figure*}
For a first investigation, three choices of the exponent~\f{m} in the defect energy approach are considered. For each choice, a corresponding normalization constant~\f{g_0} is determined by simulations such that the final nominal stress value at the final deformation strain is identical for the choices of~\f{m} (and~\f{g_0}, respectively). Consequently, for all considered exponents, the plasticity carried by the simulation specimen is identical after the final time step. The choices for~\f{m} and the corresponding normalization constants considered are shown in \tabref{tab:m_values}.
 \begin{table}[htbp]
\renewcommand{\arraystretch}{1.5}
\TC{Used defect energy exponents~\f{m} and corresponding normalization constants~\f{g_0} \rtwoC{with respective internal length scales~\f{1/g_0}}}
\label{tab:m_values}
\scriptsize
\centering
\begin{tabular}{||cccc||}
\hline
\f{m}&\f{2.0}&\f{1.5}&\f{1.1}\\\hline
\f{g_0} in 1/$\upmu$m&$8.25$&$43.5$&$450.21$\\
\rtwoC{\f{1/g_0} in $\upmu$m}&\rtwoC{$1.2\times10^{-1}$}&\rtwoC{$2.3\times10^{-2}$}&\rtwoC{$2.2\times10^{-3}$}\\\hline
\end{tabular}
\end{table}
The behavior of the model can then be compared for the cases of a quadratic defect energy, a defect energy that is closer to being linear, and a defect energy exponent that is in the range between these two cases. It is remarked that the employment of a defect energy with exponent~\f{m=1.0} would require further regularization due to the occurring jump in the gradient stresses at the beginning of the deformation. Instead, a value of~\f{m} is investigated that is sufficiently higher to not necessitate additional regularization. The stress-strain curves that are obtained after this procedure are depicted in~\figref{fig:zeta_vs_m} (left). While the quadratic defect energy leads to a linear hardening behavior, choices of lower exponents~\f{m}, however, reduce the slope of the (in general non-linear) hardening observed.
\begin{figure*}[htbp]
\centering
\includegraphics[width=\linewidth]{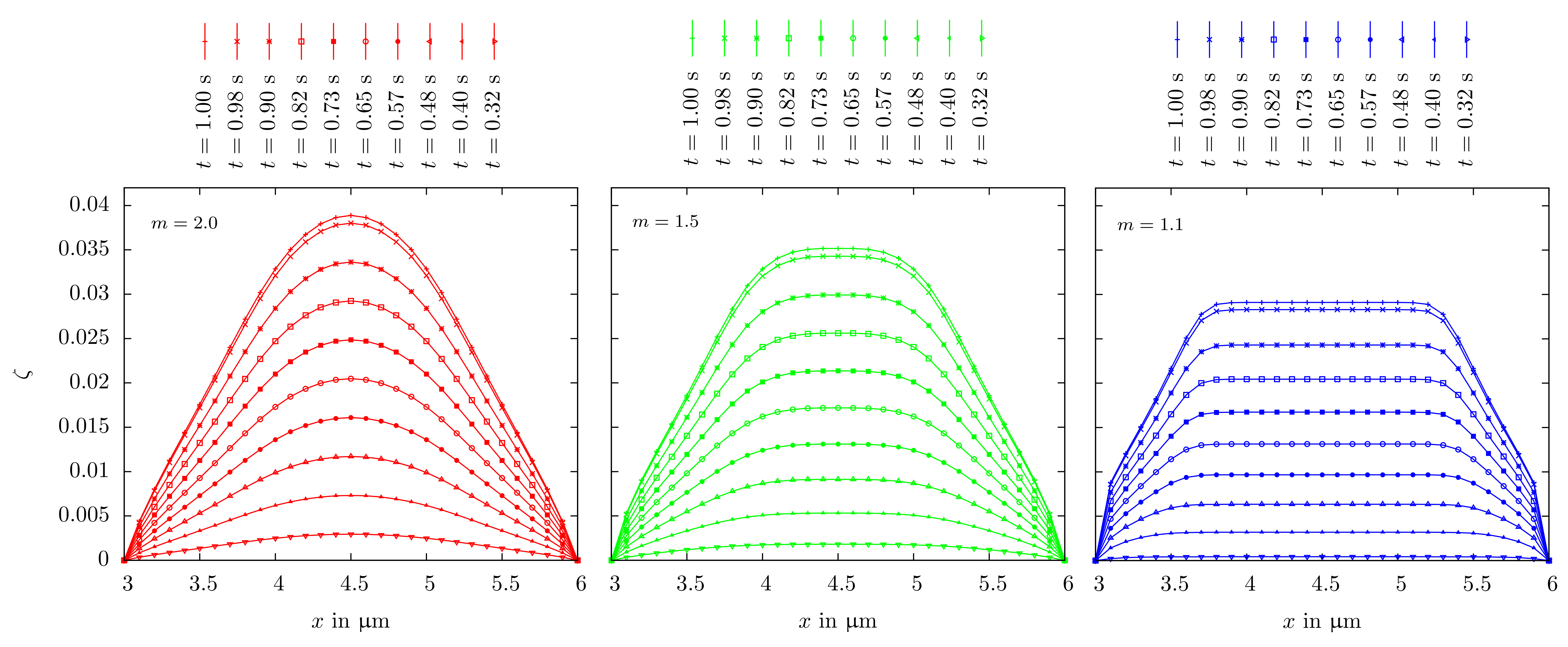}
\caption{Evolution of micromorphic variable~\f{\zeta} in the central grain of tricrystal with elastic bounding grains.}
\label{fig:zeta_vs_m_for_all}
\end{figure*}
The field distributions of the micromorphic variable~\f{\zeta} along the central line segment in the elastic-plastic grain are depicted for the three choices of~\f{m} in~\figref{fig:zeta_vs_m} (right). It can be seen there, that the common choice for the exponent of~\f{m=2.0} yields a hyperbolic profile for~\f{\zeta} in the central grain. A choice closer to~\f{m=1}, however, leads to a more plateau-like distributed~\f{\zeta} in the central grain. The gradients close to the grain boundary rise with decreasing~\f{m}. The full evolution of the \f{\zeta}-profiles is, furthermore, shown in \figref{fig:zeta_vs_m_for_all}. For the quadratic defect energy, a hyperbolic distribution of~\f{\zeta} develops. In contrast, the lower choices of~\f{m} lead to the initiation of a region with constant~\f{\zeta} that proceeds throughout the evolution. The lower the choice of~\f{m}, the more pronounced is this region. Due to the steeper gradients developing for these~\f{m}, the maximum value that~\f{\zeta} achieves is significantly less than for~\f{m=2.0}. The same ``amount'' of plasticity is carried by the central grain, however, at the end of the last time step in each case.
\subsubsection{Size effects for tricrystal with elastic bounding grains}
\begin{figure}[htbp]
\centering
\includegraphics[width=\linewidth]{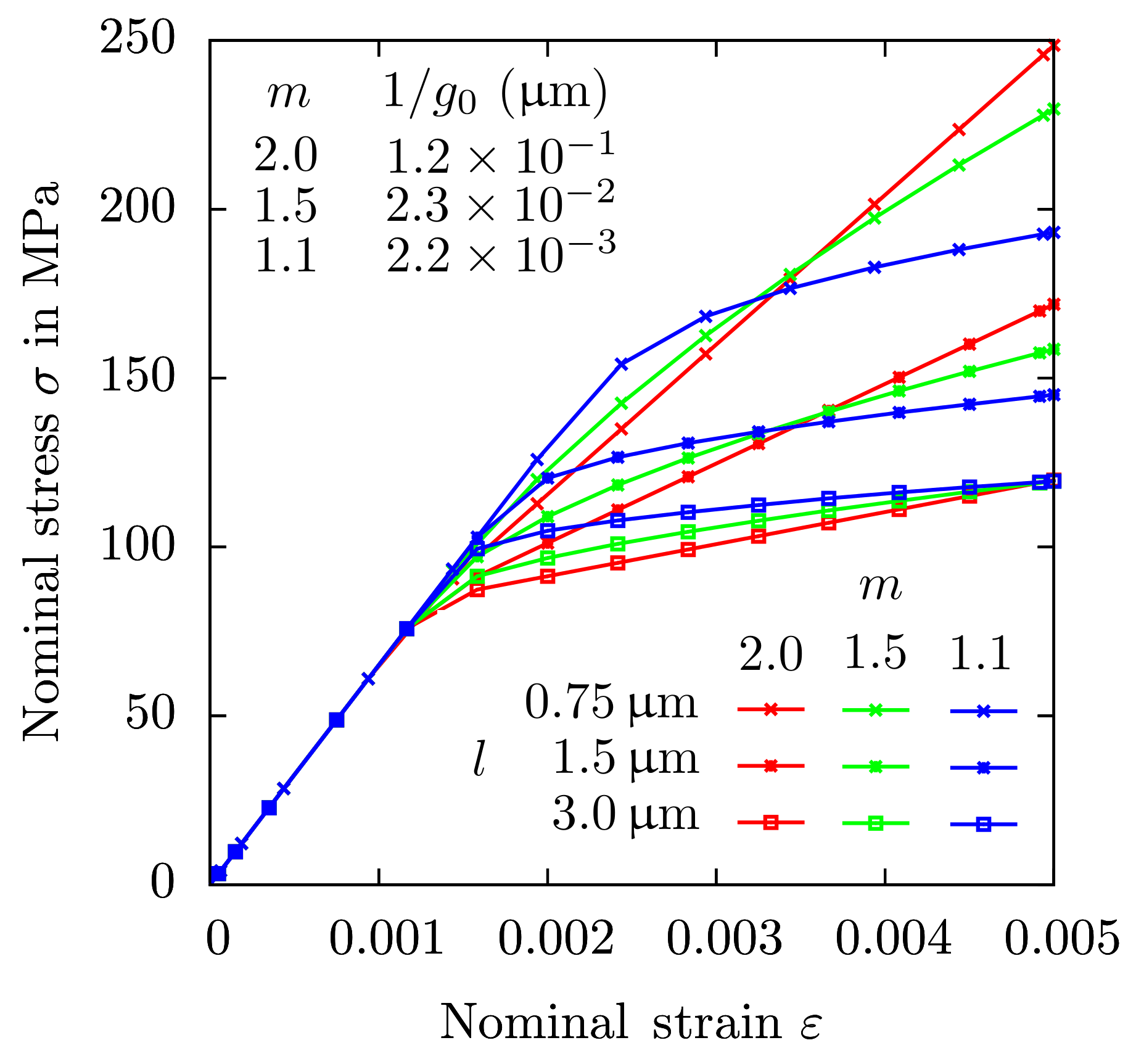}
\caption{\rtwoC{Stress-strain curves for tricrystal with elastic bounding grains for different grain sizes~\f{l} and defect energy exponents~\f{m}.}}
\label{fig:size_effects}
\end{figure}

\begin{figure}[htbp]
\centering
\includegraphics[width=\linewidth]{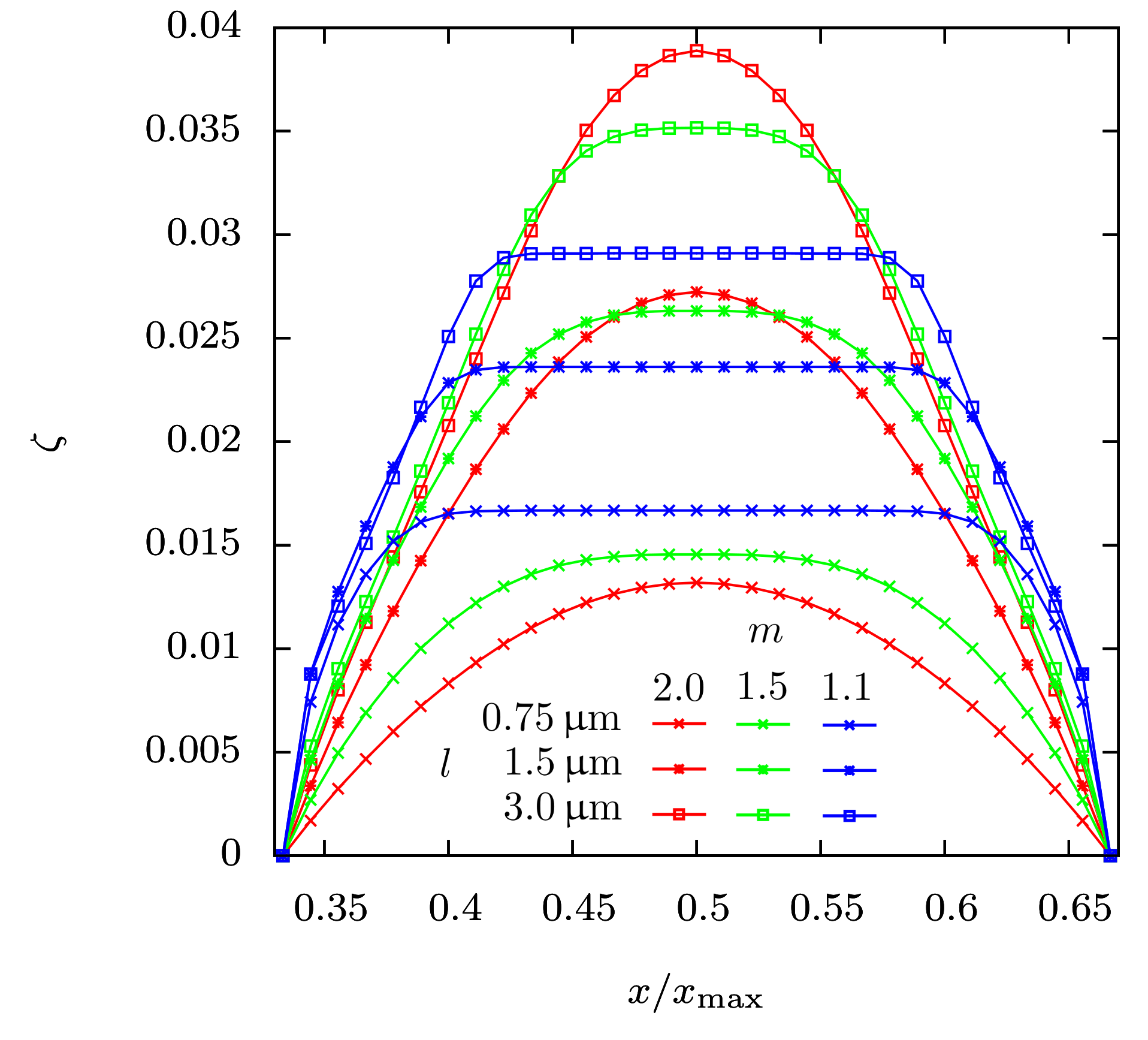}
\caption{$\zeta$-distributions in the central grain along the $x$-axis plotted over normalized coordinate $x/x_{\rm max}$ at the final time step for different grain sizes~\f{l} and defect energy exponents~\f{m}.}
\label{fig:size_effects_zeta}
\end{figure}

In addition, the size effects occurring for the different defect energy exponents are investigated with the aforementioned tricrystal. The same model parameters are considered as in the previous example, the length~$l$ of the cubic grains is, however, varied. In \figref{fig:size_effects}, the different size effect behavior is shown for three grain sizes~\f{l}.
\rtwoC{Smaller grains respond stiffer for all three values of~\f{m}, and the magnitude of the size effects is clearly influenced by the choice of the defect energy exponent and the corresponding internal length scales, which are also indicated in \figref{fig:size_effects}. It is remarked that the model behavior is influenced both by the choice of the defect energy exponent and the internal length scale. An increase in the overall yield strength becomes more pronounced for smaller~\f{m}-values. Both phenomena are discussed, in more detail, in \secref{sec:anal}. For constant defect energy exponent and length scale, the tricrystal responds stiffer with decreasing grain size~\f{l}.} A higher magnitude of the size effect is observed for the quadratic choice, compared to the smaller choices of the defect energy exponent. The smaller choices of~$m$, however, show a less pronounced size effect. In addition, the corresponding distributions of the micromorphic variable~\f{\zeta} are plotted for the final time step of each simulation in \figref{fig:size_effects_zeta}. It can be observed, there, that the size of the grains affects the distribution of~\f{\zeta}, significantly. Larger grains tend to show shorter plateaus of~\f{\zeta} and larger values of~\f{\zeta} in the center of the grain.
The classic size effect of ``smaller responding stronger'' can be observed in both \figref{fig:size_effects} and \figref{fig:size_effects_zeta}. In detail this means that the smaller the grain size, the smaller is the ''amount`` of plasticity carried by the central grain, see \figref{fig:size_effects_zeta}.
\subsubsection{Tricrystal with elastic-plastic grains}
As an additional numerical example, a tricrystal consisting of cubic grains behaving all fully elastic-plastic is investigated. This shall demonstrate the capability of the employed defect energy approach to be used in fully elastic-plastic three-dimensional multicrystalline simulations.
\begin{figure}[htbp]
\centering
\includegraphics[width=0.9\linewidth]{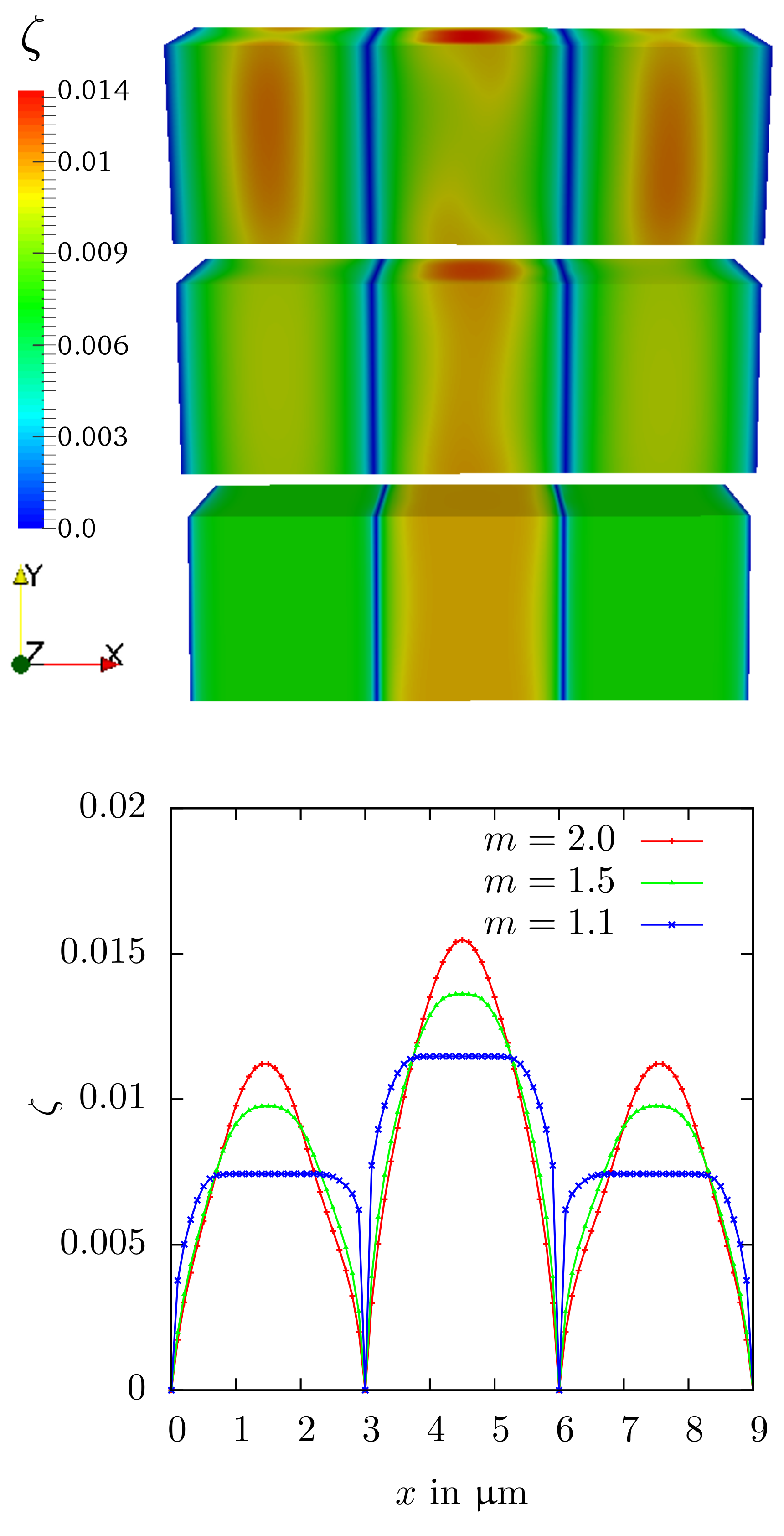}
\caption{Field distributions of~\f{\zeta} for tricrystal with elastic-plastic grains (top), and $\zeta$-distribution in the central grain along the $x$-axis (bottom) for defect energy exponents~\f{m}.}
\label{fig:zeta_vs_m_plplpl}
\end{figure}
The material parameters are chosen to be the same as in the previous numerical examples. However, the crystal orientation of the central grain is altered by a rotation of 6$^{\circ}$ around the $z$-axis (see \figref{fig:trikristall}). The resulting field distributions of~\f{\zeta} are shown in \figref{fig:zeta_vs_m_plplpl} (top) for all three choices of~\f{m}. Clearly, the width of the gradient affected zone close to the grain boundaries at~\f{x=3\,\upmu}m, and at~\f{x=6\,\upmu}m, respectively, is influenced by the defect energy exponent~\f{m}. Lower choices lead to steeper gradients that are distributed over smaller regions. This can also be seen in \figref{fig:zeta_vs_m_plplpl} (bottom), where it can be observed, as well, that the different orientations of the grains lead to differing gradients of~\f{\zeta}. For~\f{m=1.1} this becomes especially apparent with a very sharp transition in the \f{\zeta}-distribution, from the plateau in the middle of each bounding grain towards the grain boundaries.
\section{Exact Solution of a Laminate for Single Slip}\label{sec:anal}
\begin{figure*}[htbp]
\centering
\includegraphics[width=0.95\linewidth]{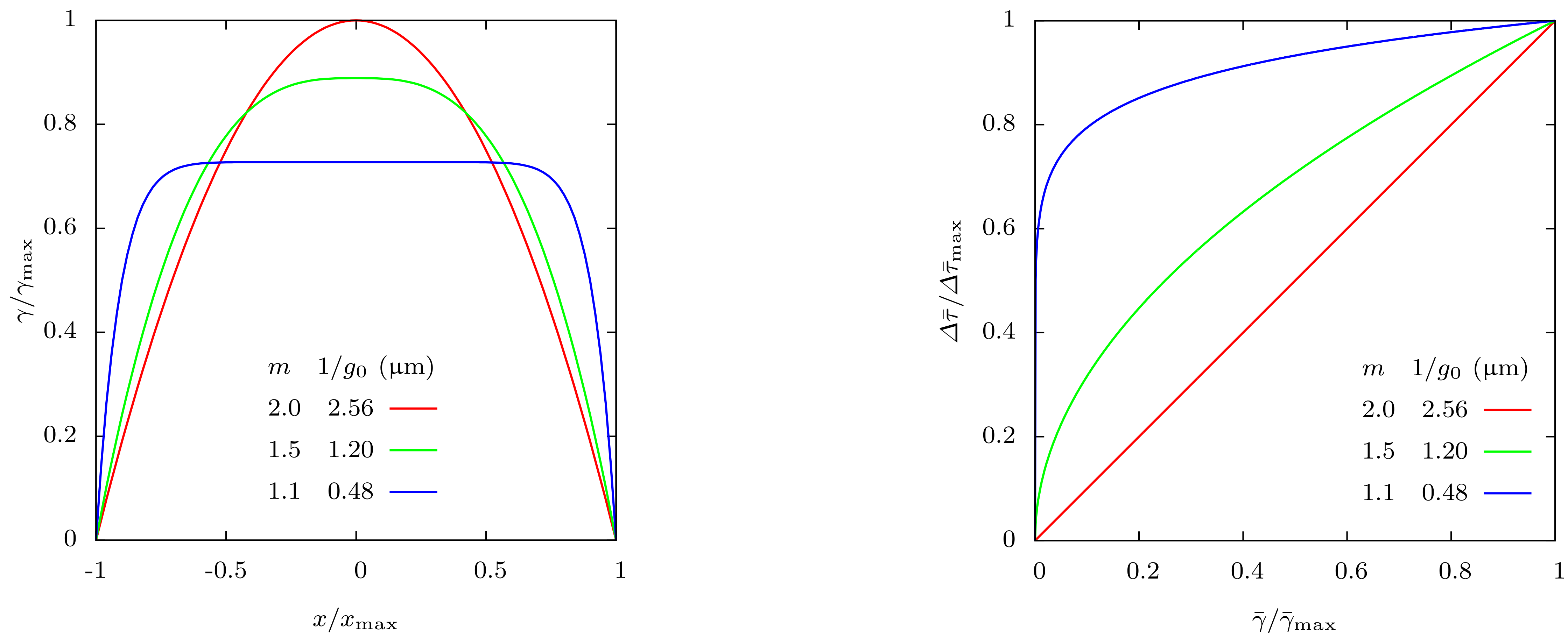}
\caption{\rtwoC{Normalized $\gamma$-distribution of analytical solution (left) for single slip simple shear, plotted along the normalized coordinate. Identical~\f{\bar\gamma} for all three values of~\f{m}.} Normalized average shear stress difference (right), plotted in the plastic range over normalized average plastic slip $\bar\gamma$.}
\label{fig:gamma_analytical}
\end{figure*}

\begin{figure*}[htbp]
\centering
\includegraphics[width=0.95\linewidth]{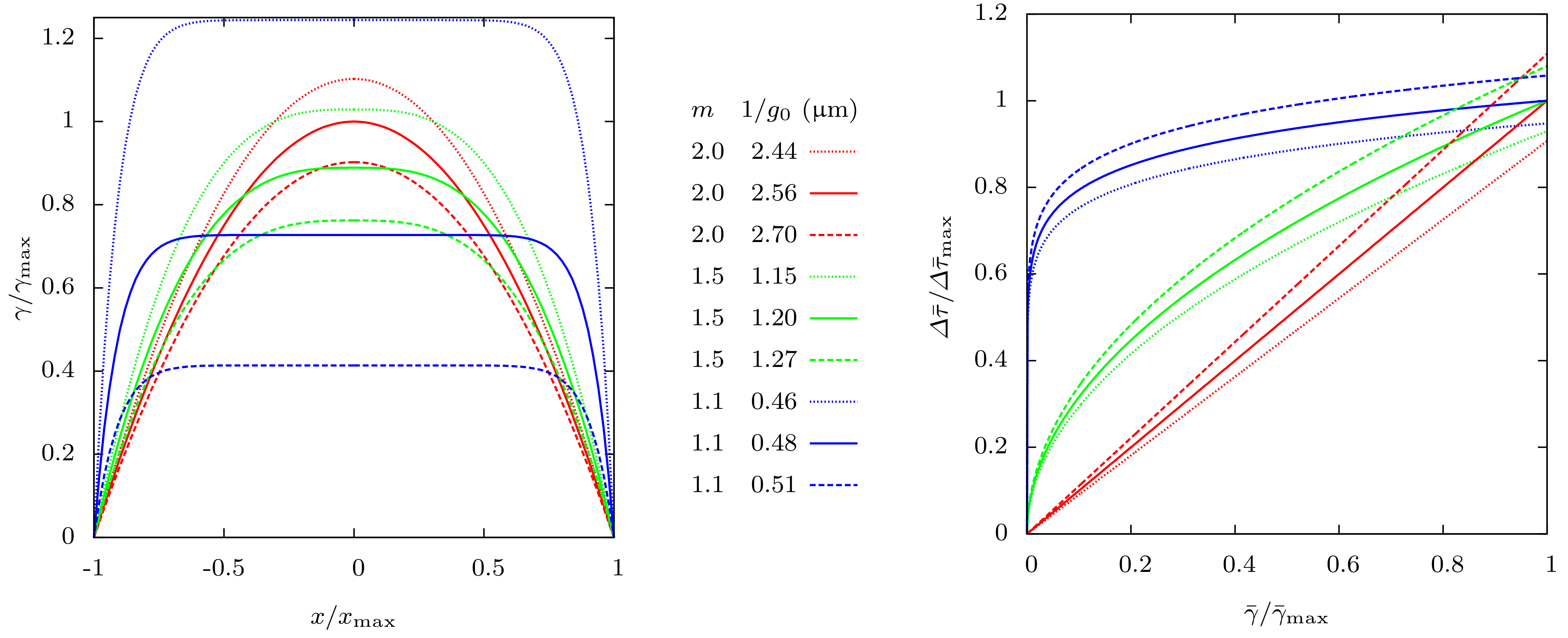}
\caption{\rtwoC{Normalized $\gamma$-distribution of analytical solution (left) for single slip simple shear, plotted along the normalized coordinate. The internal length scale is varied as indicated. Normalized average shear stress difference (right), plotted in the plastic range over normalized average plastic slip $\bar\gamma$.}}
\label{fig:gamma_analytical2}
\end{figure*}

In order to confirm the distributions of the micromorphic variable~\f{\zeta} obtained with the \rtwoC{power-law} defect energy, as shown in \secref{sec:num}, an analytical solution is briefly presented. By \equreff{eq:gradgammam}{eq:bulk_dissipation}, the gradient stress is given by
\EE
\fxi=\pd{\wgc}{\ncaq}=m\frac{W_0}{g_0}\left(\frac{\ncaqa}{g_0}\right)^{m-1}\frac{\ncaq}{\ncaqa}.
\label{eq:xi_analytical}
\Ee
In the following, single slip for simple shear is considered in the ideal case of coinciding micromorphic variable and equivalent plastic strain, i.e., \f{\zeta=\gamma_{\rm eq}=\gamma}, and~\f{\gamma=\gamma(x)}. Consequently, \f{\nabla\gamma\sim\fe_x} and the $x$-component of~\f{\fxi} reads
\EE
\label{eq:xi27}
\fxi\cdot\fe_x=\xi_x=m\frac{W_0}{g_0}\left(\frac{\nggaqa}{g_0}\right)^{m-1}.
\Ee
The term~\f{\div{\fxi}} in \eqreff{eq:tau_xi} reduces to
\EE
\div{\fxi}=\td{\xi_x}{x},
\Ee
which, in combination with \eqreff{eq:xi27} gives
\EE
\label{eq:xi5}
\td{\xi_x}{x}=(m-1)m\frac{W_0}{g_0^2}\left(\frac{\nggaqa}{g_0}\right)^{m-2}\td{\nggaqa}{x},
\Ee
where
\EE
\td{\nggaqa}{x}={\rm sg}\left(\nggaq\right)\frac{\rm d^2 \gamma}{{\rm d}x{\rm d}x}.
\Ee
For the stationary single slip case~\f{\tau^{\rm d}_\alpha=\tau^{\rm d}=\tau^{\rm C}_0}, \f{\tau_\alpha=\tau}, and, thus, \eqreff{eq:tau_xi} yields
\EE
\label{eq:xi6}
\td{\xi_x}{x}=\tau^{\rm C}_0-\tau=-\Delta\tau,
\Ee
where~\f{\Delta\tau} denotes the difference between the resolved shear stress and the critical shear stress. Combining \equreff{eq:xi5}{eq:xi6} results in
\EE
-\Delta\tau=(m-1)m\frac{W_0}{g_0^2}\left(\frac{\nggaqa}{g_0}\right)^{m-2}\td{\nggaqa}{x},
\Ee
with \f{x_{\rm min}\leq 0 \leq x_{\rm max}} and~\f{x_{\rm min}=-x_{\rm max}}. A solution for~\f{\gamma(x)} is obtained, for example, by considering \f{{\rm sg}\left({\rm d}\gamma / {\rm d}x\right)=+1} for~\f{-x_{\rm max}\leq x\leq 0} in conjunction with the BCs~\f{\tdf{\gamma}{x}(x=0)=0}, and~\f{\gamma(x=-x_{\rm max})=0}. The solution is
\EE
\begin{split}
 \gamma(x)=\frac{W_0(m-1)}{\Delta\tau}\left(\frac{\Delta\tau g_0}{W_0 m}\right)^\frac{m}{m-1}\left(-\left|x\right|^\frac{m}{m-1}\right.\eqb\left.+x_{\rm max}^\frac{m}{m-1}\right).
\end{split}
\Ee
Analogously, the solution for~\f{0\leq x \leq x_{\rm max}} can be obtained by considering~\f{{\rm sg}\left({\rm d}\gamma / {\rm d}x\right)=-1}, and the BCs \f{\tdf{\gamma}{x}(x=0)=0}, and~\f{\gamma(x=x_{\rm max})=0}, respectively.\\
In addition to the~\f{\gamma(x)}-distributions, the shear stress evolution in dependence of the average plastic slip, \f{\bar\gamma}, is derived. Therefore, the following average is considered
\begin{equation}
 \bar\gamma=\frac{1}{\Delta x}\int\limits_{x_{\rm min}}^0\gamma(x){\rm d} x,
\label{eq:bargamma}
\end{equation}
on the interval~\f{x\in[x_{\rm min},0]} with~\f{x_{\rm min}=-x_{\rm max}} and~\f{\Delta x=-x_{\rm min}}. The evaluation of \eqreff{eq:bargamma} considering an average shear stress difference of~\f{\Delta\tau=\Delta\bar\tau} leads to
\begin{equation}
 \bar\gamma=\frac{W_0(m-1)}{\Delta\bar\tau}\left(\frac{\Delta\bar\tau g_0}{W_0 m}\right)^\frac{m}{m-1}\frac{m}{2m-1}x_{\rm max}^\frac{2m-1}{m-1}.
\label{eq:bargamma2}
\end{equation}
Solving \eqreff{eq:bargamma2} for~\f{\Delta\bar\tau} yields
\begin{equation}
\Delta\bar\tau(\bar\gamma)=\bar\tau-\tau^{\rm C}_0=\frac{W_0m}{l^{2m-1}{g_0}^m}\left(\frac{2m-1}{m-1}\bar\gamma\right)^{m-1}.
\label{eq:bartau3}
\end{equation}
\rtwoC{In order to compare shear distributions for different defect energy exponents, it is assumed that the same average shear~\f{\bar\gamma=0.01} is present for the different defect energy exponents~\f{m}. Then, the corresponding values of the dimensionality constants~\f{g_0} for different~\f{m} can be obtained from \eqreff{eq:bargamma2}. This gives the values indicated in Table~\ref{tab:g0_analytical}.}
 \begin{table}[htbp]
\renewcommand{\arraystretch}{1.5}
\TC{\rtwoC{Different defect energy exponents~\f{m} and corresponding normalization constants~\f{g_0} with respective internal length scales~\f{1/g_0}, values of~\f{g_0} obtained from \eqreff{eq:bargamma2} under the assumption that~\f{\bar\gamma=0.01} for all~\f{m}}}
\label{tab:g0_analytical}
\scriptsize
\centering
\begin{tabular}{||cccc||}
\hline
\f{m}&\f{2.0}&\f{1.5}&\f{1.1}\\\hline
\rtwoC{\f{g_0} in 1/$\upmu$m}&\rtwoC{$0.39$}&\rtwoC{$0.83$}&\rtwoC{$2.07$}\\
\rtwoC{\f{1/g_0} in $\upmu$m}&\rtwoC{$2.56$}&\rtwoC{$1.2$}&\rtwoC{$0.48$}\\\hline
\end{tabular}
\end{table}
In \figref{fig:gamma_analytical} (left), the \rtwoC{normalized} distributions of~\f{\gamma(x)} are depicted for different defect energy exponents~\f{m} and assuming a shear stress difference of~\f{\Delta\tau=5}~MPa. \rtwoC{All other material parameters used are identical to the parameters in the preceding section.}
It can be observed that the smaller the defect energy exponent~\f{m} \rtwoC{and the internal length scale~\f{1/g_0}}, the higher are the gradients of~\f{\gamma(x)} close to the boundaries. A quadratic defect energy leads to a hyperbolic distribution~\f{\gamma(x)}. For choices of~\f{m} closer to one, however, more plateau-like distributions~\f{\gamma(x)} are obtained.
The resulting evolution of the average shear stress difference~\f{\Delta\bar\tau} is plotted for the plastic range in \figref{fig:gamma_analytical} (right). It can be seen, there, that a quadratic defect energy leads to uniform linear hardening, while choices of~\f{m} closer to one give \rtwoC{highly} non-linear, \rtwoC{at the onset of plasticity more pronounced}, but subsequent less hardening, compared to~\f{m=2.0}.\\
\rtwoC{If, additionally, the case \f{m=1.0} is considered, \eqreff{eq:bartau3} reduces to
\begin{equation}
\bar\tau(\bar\gamma)-\tau^{\rm C}_0=\frac{W_0}{lg_0}\Rightarrow\bar\tau(\bar\gamma)=\frac{W_0}{lg_0}+\tau^{\rm C}_0.
\end{equation}
For this case, the influence of the grain size~\f{l}, and the dimensionality constant~\f{g_0}, on the average shear stress become clear. The average shear stress~\f{\bar\tau} scales inversely linear with both, \f{l} and \f{g_0}. For exponents \f{m>1}, however, these dependencies are non-linear, see \eqreff{eq:bartau3}. This explains the increase in yield stress occurring for smaller values of~\f{m} in conjunction with smaller grain sizes~\f{l} that can be observed in the numerical results, e.g., in \figref{fig:size_effects}.}\\
\rtwoC{In the following, the influence of the defect energy exponent~\f{m} is separated from the influence of the internal length scale~\f{1/g_0} for the illustrative example above. A variation of the values of the dimensionality constant~\f{g_0} from Table~\ref{tab:g0_analytical} by~\f{\pm5}\% is performed. In \figref{fig:gamma_analytical2}, the resulting normalized average shear distributions (left) and the resulting normalized average shear stress difference courses (right) are depicted.}
\rtwoC{The courses from \figref{fig:gamma_analytical} are indicated as a reference by solid lines. From \figref{fig:gamma_analytical2}~(left), it becomes evident that the resulting shear is highly non-linear in the variations of~\f{m} and~\f{g_0}, respectively. Changes in the defect energy exponent~\f{m} further amplify variations in the gradients of~\f{\gamma} when changing the internal length scale~\f{1/g_0}. For constant~\f{m}, an increase in~\f{1/g_0} leads to more pronounced hardening behavior, see \figref{fig:gamma_analytical2}~(right), consistent with less plastic shear (see \figref{fig:gamma_analytical2}~(left)). Changing the defect exponent~\f{m}, however, gives rise to substantially different shapes of the respective hardening courses, see \figref{fig:gamma_analytical2}~(right).}
\section{Discussion of results}
The \rtwoC{power-law} defect energy approach presented has been regularized to avoid the issue of vanishing gradients in the mathematical formulation. It is shown in the present work that the regularization parameter used
 has to be chosen sufficiently small in order to be of negligible influence on the mechanical response in the gradient plasticity simulations. The choice of the defect energy exponent~\f{m} has several influences on the observed mechanical response. First of all, the resulting hardening stemming from the defect energy is influenced by the exponent choice. Furthermore, the gradients of the employed micromorphic variable close to grain boundaries are affected by the choice of~\f{m}. Smaller choices of~\f{m} lead to larger gradients. Such changes in the distribution of plastic strain can also be achieved by modifying the energetic length scale employed, see \cite{voyiadjis2014theory} and also \cite{yalcinkaya2011deformation}. The magnitude of size effects is significantly affected by the defect energy exponent choice. Smaller choices of~\f{m} lead to less pronounced size effects. It should be noted, however, that besides the energetic influences on the magnitude of size effects, also dissipative effects have been investigated in the past, e.g., by \cite{bardella2010size}. \rtwoC{For a fully elastic-plastic tricrystal ensemble of single-crystals}, the benefit of the ability to use different energy exponents~\f{m} in the model becomes especially apparent. With smaller~\f{m}, larger gradients and, thus, larger pile-ups can be modeled close to (grain) boundaries, compared to larger values of~\f{m}. The analytical solution presented confirms the influence of the choice of the defect energy exponent~\f{m} on the stress-strain curves as well as on the distribution of plastic slip and its gradients, respectively (compare \figref{fig:zeta_vs_m} and \figref{fig:gamma_analytical}). \rtwoC{In addition, the influence of the internal length scale has been shown exemplary for the analytical solution. The resulting model behavior is influenced by both the defect energy exponent and the internal length scale, see \figref{fig:gamma_analytical2}.}

\section{Conclusion}
The generalized defect energy approach of the gradient plasticity model presented in this work allows one to model different types of distributions of the micromorphic field variable~\f{\zeta} and its gradient, respectively. This can be performed by changing the exponent of the defect energy and the associated normalization constant. An analytical solution for single slip \revC{supports the numerically obtained results.} \rtwoC{It is used to show the influence of the defect energy exponent contrasted to the influence of the normalization constant. The resulting mechanical response is non-linear in the normalization constant due to the power-law defect energy approach. On one hand, changes in the exponent alter the shape of the spatial shear profiles and the shape of the work-hardening courses. On the other hand, changes of the normalization constant lead to different hardening and shear profiles for constant exponent but the qualitative shape of both the shear profiles and the hardening courses is maintained.} The used defect energy approach can be utilized \revC{on the continuum scale} to investigate the choice of the exponent in the modeling of phenomena like pile-ups \revC{of dislocations} close to grain boundaries. Furthermore, it is demonstrated in the work at hand that different magnitudes of size effects can be investigated with the underlying approach, depending on the defect energy exponent \rtwoC{and the internal length scale}. A regularization technique to resolve the issue of vanishing gradients is presented. An investigation of the choice of the regularization parameter shows no significant dependence of the model response on this parameter, if it is chosen sufficiently small. The employed regularized defect energy approach seems to be promising for future investigations with the gradient plasticity model at hand.
\appendix
\section{Derivation of field equations} \label{app:field_equations}
Application of the principle of virtual power~\f{\delta\cP_{\rm int}=\delta\cP_{\rm ext}}, i.e., letting the right-hand side of~\eqreff{eq:delta_p_int} being equal to the right-hand side of~\eqreff{eq:delta_p_ext}, yields
\EE
\begin{split}
\int\limits_\cB \left(\fsigma\cdot\delta\dot\feps+\pi\delta\dcaq+\fxi\cdot\nabla\delta\dcaq\right)\d v=\int\limits_{\partial\cB_{\rm t}}\bar\ft\cdot\delta\dot\fu\d a\eqb+\int\limits_{\partial\cB_{\Xi}}\bar\Xi\,\delta\dcaq\d a.
\end{split}
\label{eq:app_delta_p_int_ext}
\Ee
%
Then, by substituting~\f{\feps=\fsym{\grad{ \fu}}} in~\eqreff{eq:app_delta_p_int_ext} and applying two forms of the divergence theorem,
\begin{equation}
\label{eq:Green_theorem}
 \int\limits_V\fA\cdot\grad\fb\d v=-\int\limits_V\div{\fA}\cdot\fb\d v + \int\limits_{\partial V}\fA\fn\cdot\fb\d a,
\end{equation}
and
\begin{equation}
\label{eq:Green_theorem_with_jumps}
 \int\limits_V\fa\cdot\grad c\d v = -\int\limits_V\div{\fa} c\d v + \int\limits_{\partial V}\lb\fa\cdot\fn\rb c\d a,
\end{equation}
neglecting any jumps, the following is obtained after regrouping of terms
\EE
\begin{split}
-\int\limits_\cB\div\fsigma\cdot\delta\dot\fu\d v+\int\limits_{\partial\cB_{\rm t}}(\fsigma\fn-\bar\ft)\cdot\delta\dot\fu\d a\eqb+\int\limits_\cB\left(\pi-\div\fxi\right)\delta\dot\zeta\d v+\int\limits_{\partial\cB_{\Xi}}(\fxi\cdot\fn-\bar\Xi)\delta\dcaq\d a=0.
\end{split}
\label{eq:app_delta_p_int_ext_zero}
\Ee
Requiring the left-hand side of~\eqreff{eq:app_delta_p_int_ext_zero} to vanish for arbitrary virtual rates~\f{\delta\dot\fu,\delta\dot\zeta} yields                     the field equations in Box~1.

\vspace{1.5cm}\newline
{\bf Acknowledgements}\\[1ex]
The authors acknowledge the support rendered by the German Research Foundation~(DFG) under Grant BO~1466/5-1. The funded project "Dislocation based Gradient Plasticity Theory" is part of the DFG Research Group 1650 "Dislocation based Plasticity".
\vspace{-0.5cm}
\bibliographystyle{myunsrt}
\bibliography{lit_abb}
%

\end{document}